\documentclass{nature_images}
\usepackage{lineno}
\usepackage{ amssymb }
\usepackage{hyperref} 
\usepackage{amsmath}
\usepackage{caption}
\hypersetup{
  colorlinks   = true, 
  urlcolor     = blue, 
  linkcolor    = blue, 
  citecolor   = blue 
}
\def\apj{Astrophys. J.}
\def\aj{Astron. J.}
\def\mnras{Mon. Not. R. Astron. Soc}

\def\aap{Astron. Astrophys.}
\def\apjs{ApJS}

\def\actaa{Acta Astronomica}

\def\zap{Zeitschrift fuer Astrophysik}

\newcommand\ion[2]{#1$ ${\scshape{#2}}}

\title{An irradiated brown-dwarf companion to an accreting white dwarf}

\author{Juan V. Hern\'andez Santisteban$^{1}$, Christian Knigge$^1$,
  Stuart~P. Littlefair$^{2}$, Rene~P. Breton$^{3,1}$,
  Vik~S.~Dhillon$^{2,4}$, Boris~T. G\"ansicke$^{5}$,
  Thomas~R. Marsh$^{5}$, Magaretha~L. Pretorius$^6$, John
  Southworth$^7$ \& Peter~H. Hauschildt$^{8}$ }

\begin{document}

\maketitle

\begin{affiliations}
 \item Department of Physics and Astronomy, University of Southampton, Southampton SO17 1BJ, UK
 \item Department of Physics and Astronomy, University of Sheffield, S3 7RH, UK
 \item The School of Physics and Astronomy, The University of Manchester, Manchester, UK 
 \item Instituto de Astrof\'isica de Canarias, E-38205 La Laguna, Santa Cruz de Tenerife, Spain
 \item Department of Physics, University of Warwick, Coventry CV4 7AL, UK
 \item Department of Physics, University of Oxford, Denys Wilkinson Building, Keble Road, OX1 3RH, Oxford, UK
 \item Astrophysics Group, Keele University, Staffordshire, ST5 5BG, UK
 \item Hamburger Sternwarte, Gojenbergsweg 112, 21029 Hamburg, Germany

\end{affiliations}

\begin{center}

\textbf{\href{http://www.nature.com/nature/journal/v533/n7603/full/nature17952.html}{doi:10.1038/nature17952}}
\end{center}
\begin{abstract}
Brown dwarfs and giant planets orbiting close to a host star are subjected
to significant irradiation that can modify the properties of their
atmospheres. In order to test the atmospheric models that are used to
describe these systems, it is necessary to obtain accurate
observational estimates of their physical properties (masses,
radii, temperatures, albedos). Interacting compact binary systems
provide a natural laboratory for studying strongly irradiated
sub-stellar objects. As the mass-losing secondary in these systems
makes a critical, but poorly understood transition from the
stellar to the sub-stellar regime, it is also
strongly irradiated by the compact
accretor. In fact, the internal and 
external energy fluxes are both expected to be comparable
in these objects, 
providing access to an unexplored irradiation regime. 
However, the atmospheric properties of such donors have so far
remained largely unknown\cite{Littlefair:2013aa}. Here, we report the
direct spectroscopic detection and characterisation of an
irradiated sub-stellar donor in an accreting white dwarf binary
system. Our near-infrared observations allow us to determine a
model-independent mass estimate for the donor of $M_2=0.055\pm0.008M_{\odot}$
and an average spectral type of ${\rm L1}\pm{\rm1}$,
supporting both theoretical predictions and model-dependent observational
constraints. Our time-resolved data also allow us to estimate the 
average irradiation-induced temperature difference between the day and
night sides on the sub-stellar donor, $\Delta {\rm T} \simeq 57$~K,
and the maximum difference between the hottest and coolest parts of
its surface, of $\Delta {\rm T}_{max} \simeq 200$~K. The observations 
are well described by a simple geometric reprocessing model with a
bolometric (Bond) albedo of $A_B < 0.54$ at the 2-$\sigma$ confidence
level, consistent with high reprocessing efficiency, but poor lateral heat
redistribution in the donor's atmosphere\cite{Barman:2004aa,Perez-Becker:2013aa}. 
\end{abstract}

Only a single brown dwarf donor to an accreting white dwarf has been
detected spectroscopically to date\cite{Littlefair:2013aa}, and no empirical
information at all is available regarding the effect of irradiation on
the donor's atmosphere.
We have therefore carried out simultaneous optical and near-infrared
time-resolved 
spectroscopy of one such system with the X-Shooter instrument\cite{Vernet:2011aa} at
the Very Large Telescope (ESO). Our target, SDSS
J143317.78+101123.3\cite{Szkody:2007aa} (J1433 hereafter), is an eclipsing system with a 
short orbital period ($P_{orb}=78.1$ min) and a likely donor mass well below the 
hydrogen-burning limit, as estimated from eclipse
modelling ($M_2 = 0.0571 \pm 0.0007 M_{\odot}$\cite{Littlefair:2006aa,Savoury:2011aa}). The exceptionally
wide wavelength coverage provided by our data set (0.35 $\mu$m - 2.5 $\mu$m) allows
us to confidently dissect the overall spectral energy distribution. This is illustrated 
in Fig. 1. Disk-formed, double-peaked
hydrogen emission lines can be seen across the entire spectral
range. The presence of broad Balmer absorption lines between 0.3 $\mu$m
and 0.5 $\mu$m shows that the white dwarf is the strongest emitter in this wave
band. However, the donor star spectrum is clearly visible in the near-infrared and 
dominates between 1.0 $\mu$m and 2.5 $\mu$m.

The clear donor signature in the near-infrared allows us to isolate the donor's
spectrum and determine its properties. We decomposed the average
spectrum by fitting the optical region 
with a white dwarf atmosphere model\cite{Hubeny:1995aa} and an
accretion disc (modelled as a simple power law). Both of these 
contributions were then extrapolated to the near-infrared and subtracted in
order to retrieve a pure donor spectrum. This donor spectrum was then
compared to an empirical brown dwarf spectral
sequence\cite{Cushing:2005aa}. The best-matching spectral type was
found to be SpT=L1$\pm$1, which agrees remarkably well with
semi-empirical evolutionary predictions\cite{Knigge:2006kx,Knigge:2011aa} and with 
previous narrow J-band estimates\cite{Littlefair:2013aa}. 

These results imply that the donor in J1433 has successfully undergone the 
transition from the stellar to the sub-stellar regime. The alternative
-- that the secondary was born as a brown dwarf -- is extremely
unlikely. First, extreme mass-ratio binary systems containing very low
mass objects ($M\lesssim 0.1$~M$_{\odot}$) are intrinsically rare
\cite{Burgasser:2007aa}. Second, the mass of the accreting
white dwarf in J1433, M$_{WD} = 0.80\pm0.07$ M$_\odot$, is exactly in line with
the mean white dwarf mass of known accreting white dwarf systems going through the standard
evolution channel\cite{Zorotovic:2011aa}, but significantly higher
than the average mass of isolated white dwarfs  ($\langle$ M$_{WD}\rangle \simeq
0.6 $ M$_\odot$\cite{Zorotovic:2011aa}) and that of any known primary
in a white dwarf-brown dwarf binary
system\cite{Nordhaus:2013aa}. Third, the orbital period of J1433 lies
squarely within the ``period minimum spike'' that coincides with the
stellar to sub-stellar transition of the donor star in systems born
with main-sequence companions\cite{Gansicke:1999aa}. By contrast,
systems born with sub-stellar donors 
would be expected to populate a wide range of orbital periods below
this spike\cite{Politano:2004aa}. However, no such systems have ever
been found. Indeed, the only accreting white dwarf with a sub-stellar donor and an
orbital period well below $P_{min}$ has been shown to be a
low-metallicity object in the Galactic halo\cite{Uthas:2011aa}, 
rather than a system born with a brown dwarf secondary.

Models of irradiated planets suggest that irradiation can increase
photospheric temperatures by an order of magnitude compared to an 
isolated object, with significant effects on the planet's radius and
atmospheric structure\cite{Arras:2006aa}. Furthermore, a long-standing
mismatch between the predicted and observed minimum period for
accreting white dwarf binary systems can be traced to a $\sim10$\%
offset between the donor radii inferred from observations and those
predicted by theoretical model
atmospheres\cite{Littlefair:2008lr}. However, such large effects 
are only possible if the irradiating flux is efficiently absorbed
in the atmosphere and redistributed across the terminator, i.e. from
the day-side to the night-side\cite{Knigge:2011aa,Ritter:2000aa}. This
is necessary since it is the blocking effect of the 
incoming radiation on the outward heat flux from the secondary's
interior that drives the irradiation-induced swelling of the donor.

Globally, the bolometric flux emitted by the irradiated donor
must always balance the sum of its intrinsic flux and the irradiation
flux it has absorbed\cite{Rucinski:1969aa}. However, in the absence
of redistribution, this balance must be maintained locally in the
atmosphere. This should produce pronounced temperature differences
between the day- and night-sides on the tidally-locked donor. Since
the system is seen close to edge-on, such differences should manifest as changes
in the apparent spectral type of the donor as a function of orbital
phase. We have tested for this by measuring the H$_2$O 1.3 $\mu$m band
index\cite{Cushing:2005aa}. Water vapour is the most sensitive absorber in irradiated 
hot Jupiters\cite{Stevenson:2014aa}, and water
band strength has been shown to correlate well with spectral type and
effective temperature\cite{Cushing:2005aa}. As shown in Fig. 2b, we
detect a spectroscopic change of $\Delta$SpT$\simeq 1$ (i.e. M9-L0), with a
maximum at phase 0.5, precisely where the day-side of the donor is
exposed.

On its own, this spectroscopic signal would be only marginally
significant, as a model without irradiation can be ruled out at only
a 2$\sigma$ confidence level.   
However, an independent irradiation signature is provided by the observed
broad-band flux variations around the orbit, which are shown in 
Fig. 2c. This figure reveals a double modulation
over one orbit, which is produced by the combination of
two effects: the distorted shape of the tidally-locked,
Roche-lobe-filling donor star (ellipsoidal variations) and the
temperature difference between its day- and night-side (irradiation
effect). In order to quantify the irradiation effect, 
we used the binary light curve synthesis code
{\scshape{icarus}}\cite{Breton:2012aa} to fit both the water-band and
broad-band modulations under the assumption of  
no heat redistribution between hemispheres. As shown
in Fig.~2, the resulting fit is acceptable and provides a plausible representation of the
temperature distribution across the surface of the donor. 
The average temperature of the day-side is $\langle T_{day}\rangle = 2401 \pm 10$~K, and
a night-side temperature of $\langle T_{night}\rangle = 2344\pm7$~K. 
These numbers are consistent with the mean donor temperature obtained
from the template-calibrated spectral type measurements. 

The observed difference between the day-side and night-side fluxes is
statistically significant and allows us to estimate the reprocessing 
efficiency of the donor. This 
efficiency is usually parametrised via the bolometric (Bond) albedo,
$A_B$, which is the fraction of the incident irradiation flux that is not
reprocessed, but reflected back into space. 
The irradiating source in the system is the hot white dwarf. We
therefore obtained an improved measurement of its temperature,
$T_{WD}=13,200\pm200$ K, by making use of ultraviolet observations
obtained by the {\sc GALEX} satellite. With this, we find $A_B < 0.54$
at 2-$\sigma$ confidence, as shown in Extended Figure 1. 
This low value implies a high reprocessing efficiency in the
atmosphere of the donor. We can also use the observed difference
between day-side and night-side fluxes to estimate the efficiency with
which radiation absorbed locally is redistributed across the donor, $\epsilon$ (see Methods). We
find a limit for the redistribution efficiency of $\epsilon < 0.54$ at 2-$\sigma$ confidence. Efficient
reprocessing, coupled with poor heat redistribution, has also been 
found in hot Jupiters in this atmospheric temperature regime
\cite{Perez-Becker:2013aa} and suggested for 
low-mass stars in accreting white dwarfs\cite{Barman:2004aa}. If
irradiation is the main cause of the larger-than-predicted donor
radii\cite{Littlefair:2008lr}, efficient heat redistribution is
required. Our finding that day-side to night-side heat transfer must
be modest therefore suggests that irradiation is unlikely to be
responsible for inflating the donors. Instead, the dominant effect of 
irradiation is simply an increase in the local temperature\cite{Barman:2004aa}. 

In most hot Jupiters, the external irradiation overwhelms any
internal heat flux. The relatively milder irradiation experienced by
the donor in J1433, coupled with the unique prospect of obtaining
detailed phase-resolved spectra and line profiles, will permit
quantitative tests of irradiated model atmospheres in a regime 
where these observables are expected to be quite sensitive to
irradiation. The system also provides an independent new benchmark for  
theoretical models that predict the albedo and heat redistribution 
efficiency in irradiated planetary atmospheres \cite{Perez-Becker:2013aa}.
Finally, the donor in J1433 is a rapid rotator ($v\sin i=131\pm46$ km
s$^{-1}$), which could have interesting and significant effects on its
atmospheric dynamics\cite{Showman:2013aa}. 


\begin{addendum}
\item  Based on observations made with ESO Telescopes at the La Silla Paranal Observatory under programme ID 085.D-0489. J.V.H.S acknowledges support via studentships from CONACyT (Mexico) and the University of Southampton, as well as research support by the Royal Astronomical Society. R.P.B. has received funding from the European Union eleventh Framework Programme under grant agreement PIIF-GA-2012-332393. BTG was supported by ERC Grant Agreement n. 320964.
 \item[Author Contributions] C.K., S.P.L., V.S.D, B.T.G, T.R.M, M.L.P. and J.S proposed and planned the observations. All data analysis was done by J.V.H.S with significant feedback from C.K., R.P.B. and S.P.L. All authors discussed the results and commented on the manuscript.
  \item[Author Information] Correspondence and requests for materials
should be addressed to J.V.H.S. \\(email: j.v.hernandez@soton.ac.uk).
 \item[Competing Interests] The authors declare that they have no competing financial interests.

\end{addendum}

\subsection{Code Availability.}
The code used to generate the model, as well as the atmosphere
templates, {\scshape{icarus}}, is available at \url{https://github.com/bretonr/Icarus}. 

\clearpage
\begin{figure}
\includegraphics[trim=0cm 0cm 0cm 0cm, clip,width=1.0\linewidth]{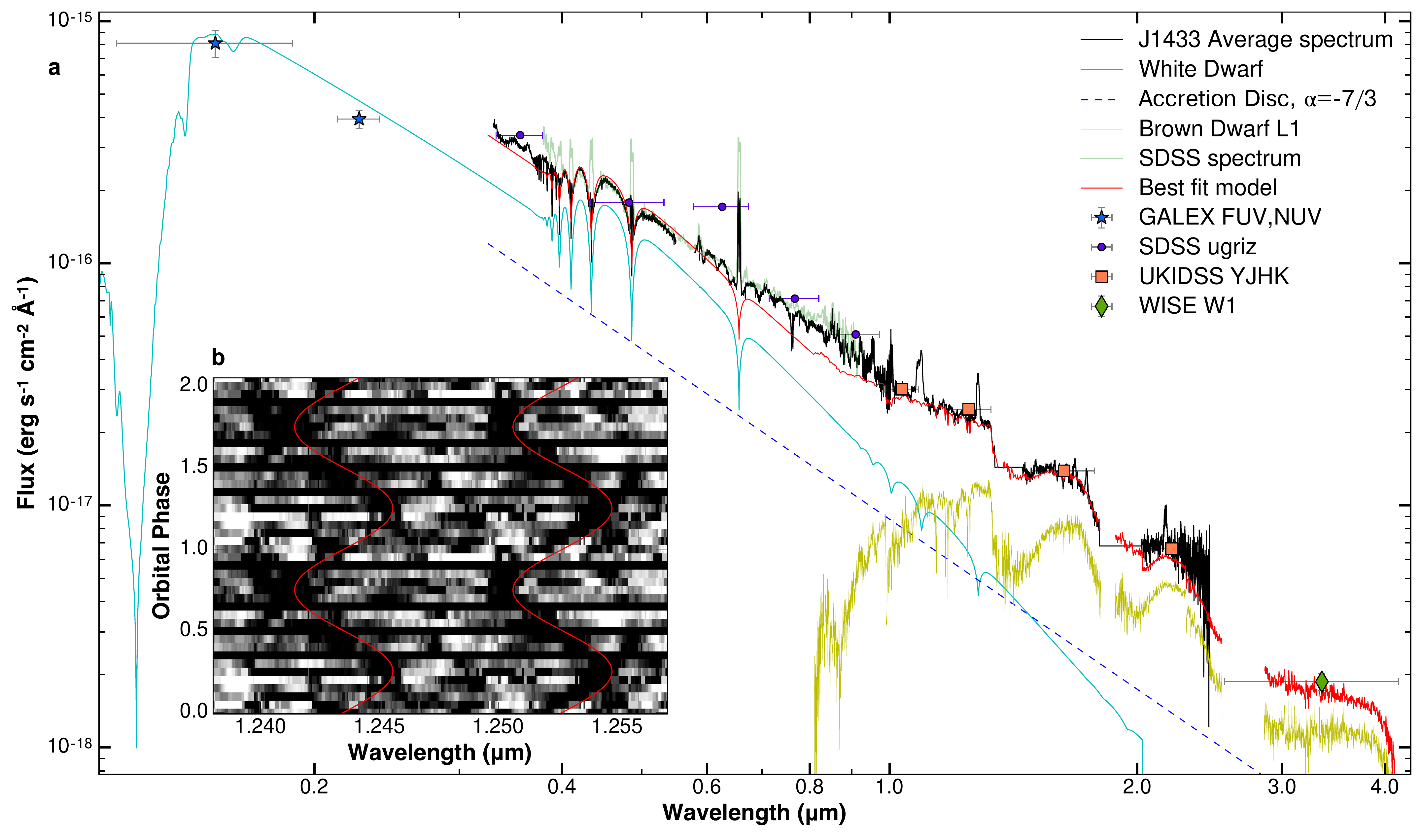} 
\captionsetup{labelformat=empty}
\caption{\textbf{Figure 1 \textbar\@ The average optical and near-infrared spectral energy distribution of
  SDSS J143317.78+101123.3.} \textbf{a}, The overall fit to the spectra
  using a white dwarf atmosphere model and brown dwarf templates was performed. The accretion
  disc contribution was described as a simple power law. We obtained a
  best fit with a spectral type of L1$\pm$1. SDSS, UKIDSS and WISE photometry
  are shown for reference. \textbf{b}, Trailed, phase-binned spectra
  around the \ion{K}{i} doublet at 1.2436 $\mu$m and 1.2528 $\mu$m. These
  features arise in the atmosphere of the donor star, enabling the
  determination of orbital parameters such as the semi-amplitude $K_2
  = 499\pm15$ km s$^{-1}$, the rotational broadening of the donor $v
  \sin i=131\pm46$ km s$^{-1}$ and systemic velocity $\gamma =
  -42\pm8$ km s$^{-1}$.} 
\end{figure}

\begin{figure}
\includegraphics[trim=0cm 0cm 0cm 1.6cm, clip,width=1.0\linewidth]{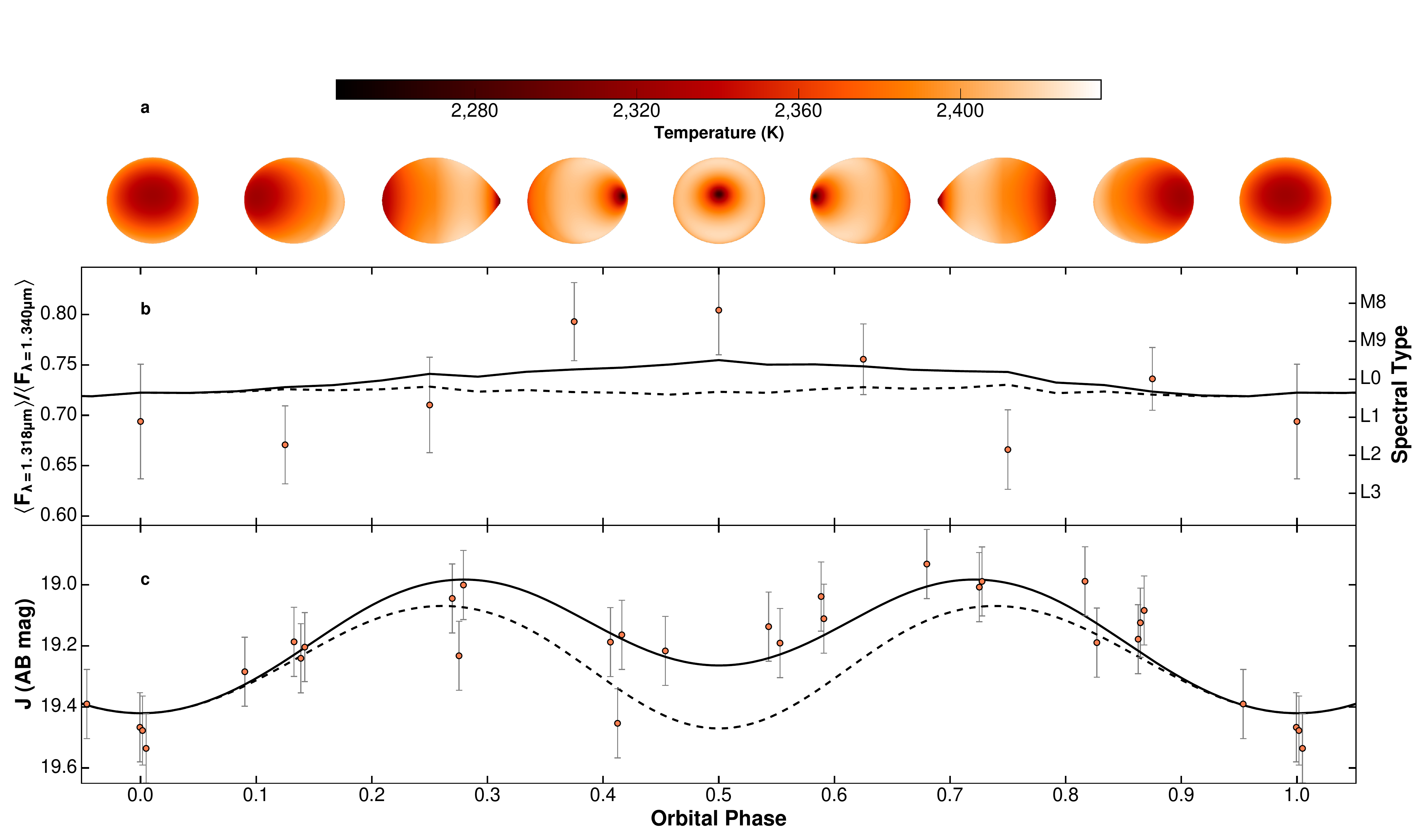} 
\captionsetup{labelformat=empty}
\caption{\textbf{Figure 2 \textbar\@ Ellipsoidal and irradiation effects associated
  with the substellar donor in SDSS J143317.78+101123.3.} We find a difference
  in spectral type between the day-side and night-side hemispheres of 
  $\Delta{\rm SpT} \simeq 1$, equivalent to $\Delta T\simeq 57$
  K with maximum temperature difference of $\simeq200$ K. \textbf{a}, Temperature
  distribution of the donor's atmosphere as a function of phase. \textbf{b}, 1.3 $\mu$m water
  band depth measurements with the best-fit model (solid line) and a model with no irradiation source (dashed line) for comparison. \textbf{c}, Timed-resolved $J$-band photometry 
  (circles)  after subtracting the white dwarf and accretion disc
  contributions. Lines represent same models as panel \textbf{b}. Error bars in both panels represent 1$\sigma$ confidence intervals.}
  \captionsetup{labelformat=empty}
\end{figure}

\clearpage

\begin{methods}
\subsection{Mass Determination}
We performed a radial velocity analysis on the the 1.243 $\mu$m and 1.252
$\mu$m \ion{K}{i} absorption lines formed in the atmosphere of the
donor. The trailed spectrum of these features is shown as the inset in
Fig. 1. We performed a simultaneous fit 
to both lines and found the best values for the semi-amplitude
$K_2=499\pm15$ km s$^{-1}$, the systemic velocity $\gamma=-42\pm8$
km s$^{-1}$ and the rotational broadening $v\sin i=131\pm46$ km s$^{-1}$. When combined with the 
observed value for the radial velocity semi-amplitude
of the WD, $K_1=34\pm4$ km s$^{-1}$ \cite{Tulloch:2009fj},
this allows us to derive a purely dynamical estimate of the mass
ratio of the system, $q = M_2/M_1 = K_1/K_2 = 0.068\pm0.008$. For this
mass ratio, the occurrence of WD eclipses alone sets a firm lower limit on
the inclination of $i > 80^\circ$\cite{Chanan:1976aa}. We can then use the mass function
of the secondary, $f(M_2) = (P_{orb}K_1^3)/(2\pi G) = M_1
\sin^3{i}/(1+q)^2$, where $G$ is the gravitational constant, to set a
purely dynamical 2$\sigma$ upper limit on 
the mass of the donor star, $M_2 < 0.071 M_{\odot}$. We can also use the
mass function to obtain a model-independent {\em estimate} of the
donor mass. In any semi-detached compact binary system, there is a
unique, single-valued family of $i-q$ pairs that produce a compact
object eclipse with a given width\cite{Chanan:1976aa}. In the case of
J1433, the WD eclipse width can be measured 
directly from published optical light
curves\cite{Littlefair:2008lr}. We can then combine the resulting
$i-q$ constraint with our 
spectroscopic mass ratio and the mass function to obtain a robust
donor mass estimate that depends only on Kepler's third law and the
well-understood geometry of Roche-lobe-filling objects. We find 
M$_2=0.055\pm0.008$ M$_{\odot}$, well below the hydrogen-burning limit.
\subsection{Water-band ratios and broad-band fluxes.}
In order to provide the cleanest measurements of the donor at any
given phase, we subtracted both white dwarf and accretion
disc contributions. The flux observed from the white dwarf was assumed
to be constant in phase, except for phase bins affected by an
eclipse. In these, we reduced the flux by the fraction of the bin
width during which the white dwarf was occulted. The accretion disc
contribution was modelled by fitting a power-law to the optical
(3200--5400 \AA\@) waveband (while masking the emission lines) and
extrapolating to the NIR. Due to the low S/N of the individual
spectra, the telluric removal created artefacts in some regions of the spectra. We
identified these regions and masked them before shifting each spectrum
to the donor's rest frame.  
We binned the data into eight orbital phase bins and measured a slightly
modified water-band index, defined as $\langle F_{\lambda=1.318 \mu m}
\rangle/\langle F_{\lambda=1.340 \mu m} \rangle$, where the fluxes are
averages over a 5~\AA\@ window. 
This index was then calibrated against brown dwarf spectral type
templates\cite{Cushing:2005aa}. Errors were estimated via 500
bootstrap copies of the median flux inside every orbital bin for all
the spectra. The broad-band flux modulation was estimated by 
integrating the donor flux over the $J$-band (after masking all
emission lines) and converting to AB magnitudes. 
\subsection{Day- and night-side temperatures, albedo and intrinsic luminosity.}
\label{sec:model}
In order to model the orbital-phase-dependent water band ratios and
broad-band flux, we constructed a grid of irradiated donor models with 
{\scshape{icarus}}\cite{Breton:2012aa}. Each of these models is
characterized by two parameters: the irradiating luminosity
absorbed by the donor, and the intrinsic luminosity it would produce in
the absence of irradiation. For each parameter pair, we use
{\scshape{icarus}} to calculate the resulting temperature distribution
across the donor and to predict the observed spectrum at each orbital
phase. These calculations assume that all of the irradiating flux
absorbed at a given point on the donor's surface is re-emitted
locally and neglect any effect of the donor's fast rotation
on the temperature distribution (which could affect the atmospheric
dynamics\cite{Showman:2013aa}). However, the distorted shape of the
Roche-lobe-filling donor, as well as limb- and gravity-darkening, are
fully taken into account in these  
calculations. 
The absorbed irradiating and intrinsic luminosities are
not directly observable, but they respectively determine the day-side
and night-side temperatures, which are. We therefore present our
results in terms of these temperatures, which we calculate by
summing over all day-side surface elements in the model, 
\begin{equation*}
\langle T_{\text{day}}\rangle   =  \left(\frac{1}{\sigma} \frac{ \sum_{j,\text{day}} F_{\text{tot}_j} }{\sum_{j,\text{day}}A_{j}}  \right)^{1/4},
\end{equation*}
and all night-side elements,
\begin{equation*}
\langle T_{\text{night}}\rangle =  \left(\frac{1}{\sigma} \frac{ \sum_{j,\text{night}} F_{\text{int},j} }{\sum_{j,\text{night}}A_{j}}  \right)^{1/4}.
\end{equation*}
Here, $F_{\text{int},j}$ is the intrinsic flux and is given by 
\begin{equation*}
F_{\text{int},j} = A_j \sigma T_{\text{int},j}^4,
\end{equation*}
where $A_j$ is the area of the surface element, and $\sigma$ is the
Stefan-Boltzmann constant. As a consequence of the distorted shape of
the donor, $T_{\text{int},j}$ --
the temperature of each surface
element in the absence of irradiation -- is not constant. Similarly,
the total flux that heats each surface element on the
day-side, $F_{\text{tot},j}$, is given  
\begin{equation*}
F_{\text{tot},j} = F_{\text{int},j} + (1-A_B)F_{\text{irr},j}, 
\end{equation*}
where $A_B$ is the bolometric (Bond) albedo of the donor. 
$F_{\text{irr},j}$ is the irradiating flux seen by this element, given by 
\begin{equation*}
F_{\text{irr},j} =  \frac{4\pi R_{\text{WD}}^2\sigma
  T_{\text{WD}}^4}{4\pi d_j^2} A_{\text{proj},j}. 
\end{equation*}
Here $d_j$ is the distance of every surface element to the white
dwarf, $T_{\text{WD}}$ is the white dwarf temperature and
$A_{\text{proj},j}$ is the projected surface intercepting the
irradiating flux. Since we have good estimates of the white dwarf
temperature and binary parameters, $F_{\text{irr},j}$ is well-constrained
observationally. Thus the {\em absorbed} irradiating flux and the
day-side temperature can equivalently be thought of as measures of the
albedo. We determine optimal values of $\langle
T_{\text{day}}\rangle$, $\langle T_{\text{night}}\rangle$, and $A_B$
by fitting the observed phase-dependent water-band and broad-band
fluxes using a Markov-Chain Monte Carlo procedure. The results are
shown in the Extended Data Figure~1. 
We find $\langle T_{\text{night}}\rangle=2344^{+11}_{-10}$ K and
$\langle T_{\text{day}}\rangle=2401^{+7}_{-7}$~K, where the errors
represent the 1$\sigma$ confidence levels. The corresponding
constraints on the albedo and intrinsic donor luminosity are $A_B <
0.54$ (at 2-$\sigma$) and $L_{int} = 3.1 \pm 0.1 \times 10^{-4} L_{\odot}$. 
In all of these models, we
adopt a white dwarf temperature of $T_{\text{WD}} = 13200\pm200$~K, based on 
our own analysis of ultraviolet data obtained by {\sc GALEX},
$i=84.36^{\circ}$ for the inclination, $d=226$ pc for the
distance, and $P_{orb}=78.106657$ min for the orbital periods, based on
previous eclipse modelling estimates\cite{Savoury:2011aa}. 
We also adopt a fixed gravity-darkening coefficient of
$\beta=0.08$\cite{Lucy:1967aa}. z

\subsection{Heat Redistribution Across the Terminator.}
\label{redist}
Our geometric reprocessing model assumes that none of the irradiating
flux is redistributed from the day-side to the night-side. This model
fits the data acceptably, suggesting that redistribution is
inefficient in the atmosphere of the donor. In order to estimate this
efficiency quantitatively, we need to adopt a specific model for
redistribution. Here, we assume that a constant fraction $0 \leq
\epsilon \leq 1$ of the irradiating flux that is absorbed at any point
on the day-side of the donor is redistributed evenly across the entire
surface. Similar models are commonly used in
studies of irradiated exo-planets \cite{Cowan:2011aa}. In the context
of this simple model, the difference between the fluxes emerging from
the day-side and the night-side is given by
\begin{equation*}
\sigma \langle T_{\text{day}}\rangle^4  - \sigma \langle T_{\text{night}}\rangle^4  = 
\left(\sigma\langle T_{\text{day}}\rangle^4  - \sigma \langle T_{\text{night}}\rangle^4\right)_{\epsilon=0} -
\epsilon(1-A_B) \frac{\sum_{j,\text{day}} F_{\text{irr}_j} A_j}{\sum_{j,\text{day}} A_j},
\end{equation*}
where $\left(\sigma\langle T_{\text{day}}\rangle^4  - \sigma \langle
T_{\text{night}}\rangle^4\right)_{\epsilon=0}$ is the flux difference
of the equivalent model without any redistribution. The modelling
described in the previous section provides us with a set of
set of self-consistent flux differences for models with $\epsilon =
0$. We can therefore transform these to the flux differences 
predicted for otherwise equivalent models with $\epsilon > 0$. Since
our earlier modelling also provides us with estimates of the
{\em observed} day-side and night-side temperatures, we can estimate
$\epsilon$ by comparing the observationally inferred flux difference
to that predicted by our transformed models with $\epsilon > 0$.
The results are shown in Extended Figure~2. We find that the efficiency
of redistribution is indeed low, $0.1 < \epsilon < 0.4$, with higher
values corresponding to lower albedos. The 2~$\sigma$ upper limit on
the redistribution efficiency (for $A_{B} = 0$) is $\epsilon < 0.54$. 


\end{methods}
\clearpage
\begin{figure}
\includegraphics[trim=0cm 0cm 0cm 0cm, clip,width=1.0\linewidth]{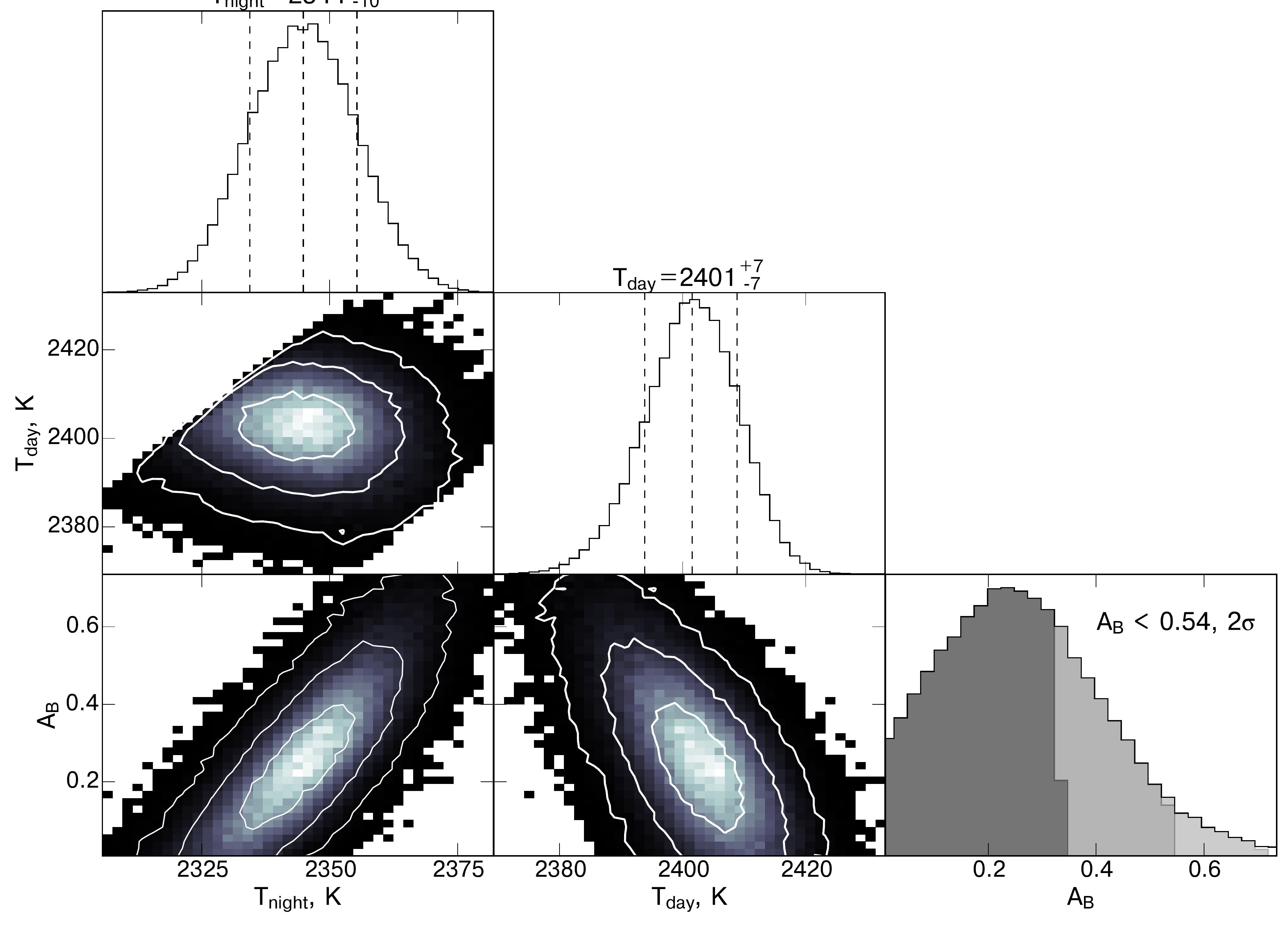} 
\captionsetup{labelformat=empty}
\caption{\textbf{Extended Data Figure 1 \textbar\@ Posterior probability distributions for the irradiation model parameters.} Colour scale contours show the joint probability for every combination of parameters. Contours represent the 1, 2 and 3-$\sigma$ levels. Marginal posterior distributions are shown as histograms with the median and 1-$\sigma$ marked as dashed lines. The A$_B$ distribution is quoted as a truncated distribution with a 2-$\sigma$ upper limit.}
\end{figure}

\begin{figure}
\includegraphics[trim=0cm 0cm 0cm 0cm, clip,width=1.0\linewidth]{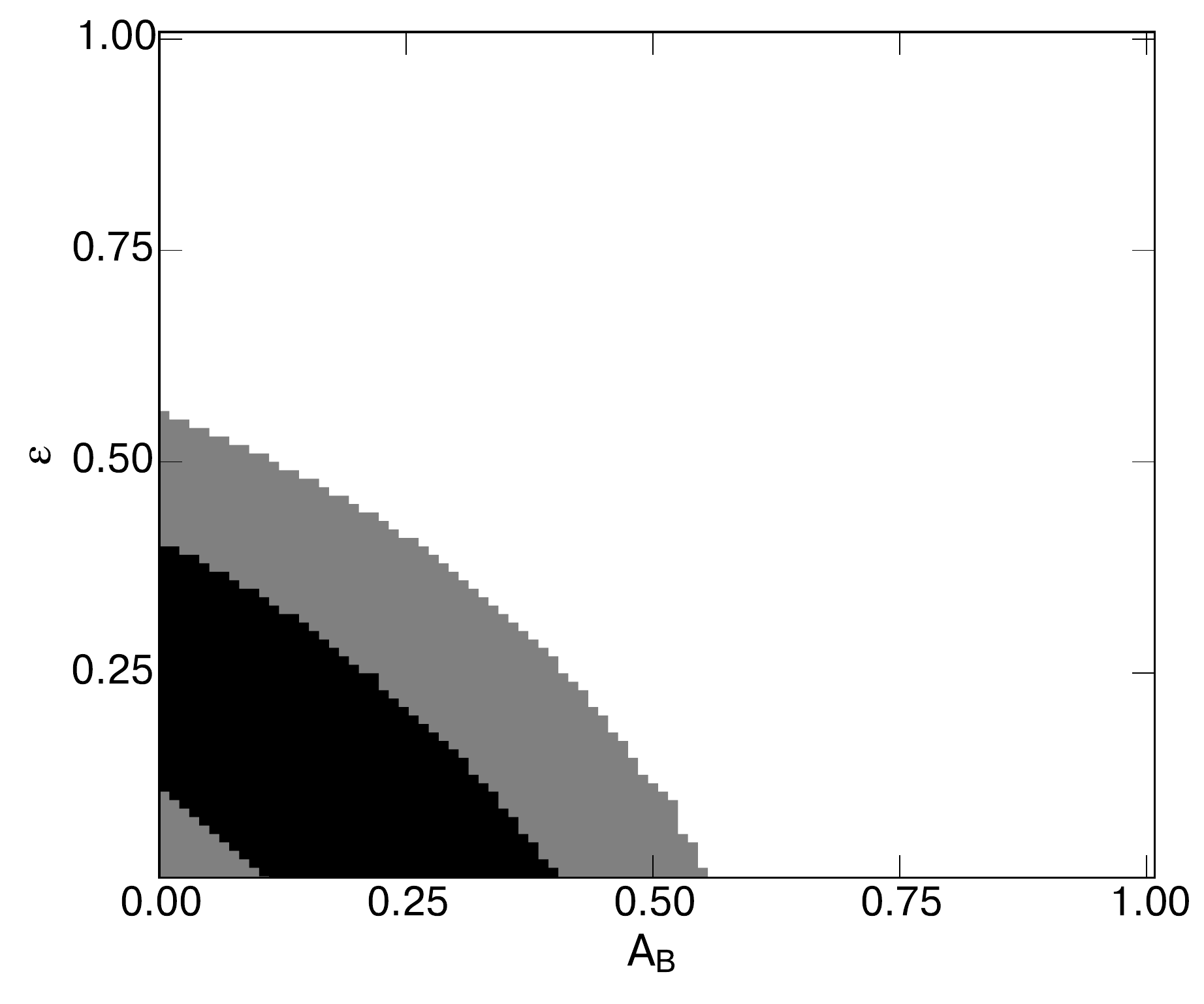} 
\captionsetup{labelformat=empty}
\caption{\textbf{Extended Data Figure 2 \textbar\@ Redistribution efficiency limits for the irradiated sub-stellar donor.} Allowed family solutions of redistribution efficiencies as a function of Bond albedo are shown. The black and grey contours represent the 1 and 2-$\sigma$ confidence levels.}
\end{figure}

\end{document}